\documentclass[twocolumn,preprintnumbers,amsmath,amssymb,superscriptaddress]{revtex4}
\usepackage{graphicx}
\usepackage{dcolumn}
\usepackage{bm}
\usepackage{soul}
\usepackage{color}
\usepackage{epstopdf}
\usepackage[version=3]{mhchem}
\usepackage{lipsum}
\usepackage[outercaption]{sidecap}
\usepackage{floatrow}
\begin{document}

\title{Theoretical and Experimental Study of Compression Effects on Structural Relaxation of Glass-Forming Liquids}% Force line breaks with \\

\author{Anh D. Phan}
\affiliation{Faculty of Materials Science and Engineering, Phenikaa Institute for Advanced Study, Phenikaa University, Hanoi 12116, Vietnam}
\email{anh.phanduc@phenikaa-uni.edu.vn}
\affiliation{Faculty of Computer Science, Artificial Intelligence Laboratory, Phenikaa University, Hanoi 12116, Vietnam}
\affiliation{Department of Nanotechnology for Sustainable Energy, School of Science and Technology, Kwansei Gakuin University, Sanda, Hyogo 669-1337, Japan}
\author{Agnieszka Jedrzejowska}
\affiliation{Institute of Physics, University of Silesia, SMCEBI, 75 Pulku Piechoty 1a, 41-500 Chorzow, Poland}
\author{Marian Paluch}
\affiliation{Institute of Physics, University of Silesia, SMCEBI, 75 Pulku Piechoty 1a, 41-500 Chorzow, Poland}
\author{Katsunori Wakabayashi}
\affiliation{Department of Nanotechnology for Sustainable Energy, School of Science and Technology, Kwansei Gakuin University, Sanda, Hyogo 669-1337, Japan}
%\author{Vu D. Lam}
%\affiliation{Institute of Materials Science, Vietnam Academy of Science and Technology, Hanoi, Viet Nam}
%\affiliation{Graduate University of Science and Technology, Vietnam Academy of Science and Technology, Hanoi, Viet Nam}
\date{\today}

\date{\today}

\begin{abstract}
We develop the elastically collective nonlinear Langevin equation theory of bulk relaxation of glass-forming liquids to investigate molecular mobility under compression conditions. The applied pressure restricts more molecular motion and therefore significantly slows-down the molecular dynamics when increasing the pressure. We quantitatively determine the temperature and pressure dependence of the structural relaxation time. To validate our model, dielectric spectroscopy experiments for three rigid and non-polymeric supramolecules are carried out at ambient and elevated pressures. The numerical results quantitatively agree with experimental data. 
\end{abstract}

\maketitle

\section{Introduction}
Investigating molecular dynamics of glass-forming liquids is one of the most intriguing topics. It has been experimentally established that the structural relaxation time ($\tau_\alpha$) reflecting the time scale for liquid structure reorganization systematically deviates from the simple Arrhenius behaviour during cooling process on approaching to the glass transition temperature, $T_g$, defined by $\tau_\alpha(T_g)=100$ s \cite{61,62}. The non-Arrhenius dependence of the structural relaxation time, $\tau_\alpha$ has universal character because it has been observed for different groups of glass-forming liquids (van der Waals and associated liquids, polymers, ionic liquids, molten metals, etc). However, degree of deviation of $\tau_\alpha$ form the Arrhenius law at $T_g$ is material dependent and is characterized by means of fragility or steepness index, $m=\left[\partial\log_{10}\tau_\alpha/\partial (T_g/T)\right]_{T=T_g}$. Consequently, the parameter $m$ was used to introduce the strong versus fragile liquid classification scheme. According to this classification strong liquids reveal temperature evolution of structural relaxation time less deviating from the Arrhenius behaviour than fragile ones. 

Much efforts have been spent in the last decades to formulate satisfactory models being able to capture and explain all experimentally observed features of structural dynamics of glass forming liquids. One of such successful approaches is the Elastically Collective Nonlinear Langevin Equation (ECNLE) theory of bulk relaxation \cite{2,6,7,10,8,42,9}. In this theory, a single molecular motion is considered as a consequence of its interactions with the nearest neighbours and molecular cooperativity outside the cage of neighbouring molecules. The treatment leads to two strongly-related but distinct barriers corresponding to local and elastically collective dynamics. Plugging these two barriers into the Kramer's theory gives the structural alpha relaxation times. To determine the temperature dependence of the structural relaxation times, Mirigian and Schweizer have used a thermal mapping, which is based on an equality between hard-sphere-fluid and experimental isothermal compressibility. From this, the ENCLE theory has successfully described the alpha relaxation event of polymers \cite{2,8}, and thermal liquids \cite{6,7,10} over 14 decades in time. However, amorphous drugs and many materials have no experimental data for the thermal mapping. It is impossible to compare ECNLE calculations with experiments. Recently, Phan and his coworkers \cite{42,9,50} proposed another density-to-temperature conversion based on the thermal expansion process to handle this issue.

The rapid cooling of liquid to obtain the glass is not the only way. Alternative method to vitrify it is squeezing (compression) \cite{63,64}. Therefore, by changing the hydrostatic pressure of liquid one can also control its molecular dynamics \cite{63,64}. Compression brings about increase in the molecular packing, in consequence, leading to increase of the structural relaxation time. Numerous experimental results \cite{63,64} show that the pressure counterpart of the Arrhenius law: 
\begin{eqnarray}
\tau_\alpha=\tau_0\exp\left(\frac{P\Delta V}{k_BT}\right),
\label{eq:1}
\end{eqnarray}
derived based on transition state theory fails to grasp pressure dependence of $\tau_\alpha$, where $\Delta V$ is activation volume, $P$ is the pressure, and $k_B$ is Boltzmann constant. The experimentally measured relaxation times are found to change with pressure much faster than predicted by Eq. (\ref{eq:1}) \cite{64,65,66,67,68}. It indicates that the activation volume is not constant but in general increases with increasing pressure on approaching to glassy state. An extension of ECNLE theory \cite{10} was introduced in 2014 to understand compression effects on the glass transition. Authors used Schweizer's the thermal mapping associated with the compressibility data measured at different pressures. However, theoretical predictions are more sensitive to pressure than experiments. Thus, it is crucial to propose a better model to determine quantitatively the pressure-dependent structural dynamics.  

The main goal of this paper is to develop the ECNLE theory in a new approach to describe the pressure dependence of $\tau_\alpha$. To validate our development, we implement new dielectric spectroscopy measurements on three different rigid and non-polymeric supramolecules at a wide range of pressures and temperatures. Then, theoretical calculations are quantitatively compared to experimental results. Theoretical limitations are clearly discussed.

\section{Theoretical Methods}
\subsection{Formulation}
To theoretically investigate the structural relaxation time of amorphous materials, these materials are described as a fluid of disconnected spheres (a hard-sphere fluid) and we formulate calculations for activation events of a single particle. The hard-sphere fluid is characterized by a particle diameter, $d$, and the number of particles per volume, $\rho$. According to the ECNLE theory \cite{2,6,7,10,8,42,9,3,4}, the dynamic free energy quantifying interactions of an arbitrary tagged particle with its nearest neighbors at temperature $T$ is

\begin{eqnarray}
\frac{F_{dyn}(r)}{k_BT} &=& \int_0^{\infty} dq\frac{ q^2d^3 \left[S(q)-1\right]^2}{12\pi\Phi\left[1+S(q)\right]}\exp\left[-\frac{q^2r^2(S(q)+1)}{6S(q)}\right]
\nonumber\\ &-&3\ln\frac{r}{d},
\label{eq:2}
\end{eqnarray}
where $\Phi = \rho\pi d^3/6$ is the volume fraction, $S(q)$ is the static structure factor, $q$ is the wavevector, $r$ is the displacement of the particle. The dynamic free energy is constructed without considering effects of rotational motions. We use the Percus-Yevick (PY) integral equation theory \cite{1} for a hard-sphere fluid to calculate $S(q)$. The PY theory defines $S(q)$ via the direct correlation function $C(q)=\left[S(q)-1 \right]/\rho S(q)$. The Fourier transform of $C(q)$ is \cite{1}
\begin{eqnarray}
C(r) &=& -\frac{(1+2\Phi)^2}{(1-\Phi)^4} + \frac{6\Phi(1+\Phi/2)^2}{(1-\Phi)^4}\frac{r}{d}\nonumber\\
&-&\frac{\Phi(1+2\Phi)^2}{2(1-\Phi)^4}\left(\frac{r}{d}\right)^3 \quad \mbox{for} \quad r \leq d \\
C(r) &=& 0 \quad \mbox{for} \quad r > d.
\end{eqnarray}

The free energy profile gives us important information for local dynamics. For $\Phi \leq 0.43$, $F_{dyn}(r)$ monotonically decreases with increasing $r$ and particles are not localized \cite{3,4,1}. In denser systems ($\Phi > 0.43$), one observes the dynamical arrest of particles within a particle cage formed by its neighbors occurs and a free-energy barrier emerges as shown in Fig. \ref{fig:3}. We determine the particle cage radius, $r_{cage}$, as a position of the first minimum in the radial distribution function, $g(r)$. The localization length ($r_L$) and the barrier position ($r_B$) are the local minimum and maximum of the dynamic free energy. The separation distance between these two positions, $\Delta r =r_B-r_L$, is a jump distance. The local energy-barrier height is calculated by $F_B=F_{dyn}(r_B)-F_{dyn}(r_L)$.

Compression effects modify motion of a single particle. Motion of a particle is governed by both nearest-neighbor interparticle interactions and applied pressure. Under a high pressure condition, when a particle is displaced by a small distance ($r \ll d$), the applied pressure acts on a volume $\Delta V(r) \approx d^2r$ and causes the mechanical work. In addition, the free volume and the molecular volume are reduced with compression. For simplification purposes, we suppose that the volume fraction is insensitive to pressure. Thus, we propose a new and simple expression for the dynamic free energy
\begin{eqnarray} 
\frac{F_{dyn}(r)}{k_BT} %&\equiv& \frac{F_{dyn}(r)}{k_BT} + P\frac{d^2r}{k_BT}, \nonumber\\
&\approx& \int_0^{\infty} dq\frac{ q^2d^3 \left[S(q)-1\right]^2}{12\pi\Phi\left[1+S(q)\right]}\exp\left[-\frac{q^2r^2(S(q)+1)}{6S(q)}\right]
\nonumber\\ &-&3\ln\frac{r}{d} + \frac{P}{k_BT/d^3}\frac{r}{d}.
\label{eq:pressure}
\end{eqnarray}

The diffusion of a particle through its cage is decided by rearrangement of particles in the first shell. The reorganization process slightly expands the particle cage and excites collective motions of other particles in surrounding medium by propagating outward radially a harmonic displacement field $u(r)$. By using Lifshitz's continuum mechanics analysis \cite{5}, the distortion field in a bulk system is analytically found to be
\begin{eqnarray} 
u(r)=\frac{\Delta r_{eff}r_{cage}^2}{r^2}, \quad {r\geq r_{cage}},
\label{eq:3}
\end{eqnarray}
where $\Delta r_{eff}$ is the cage expansion amplitude \cite{6,7}, which is
\begin{eqnarray} 
\Delta r_{eff} = \frac{3}{r_{cage}^3}\left[\frac{r_{cage}^2\Delta r^2}{32} - \frac{r_{cage}\Delta r^3}{192} + \frac{\Delta r^4}{3072} \right].
\end{eqnarray}
Since $\Delta r_{eff}$ is relatively small, particles beyond the first coordination is supposed to be harmonically oscillated with a spring constant at $K_0 = \left|\partial^2 F_{dyn}(r)/\partial r^2\right|_{r=r_L}$. Thus, the oscillation energy of the oscillator at a distance $r$ is $K_0u^2(r)/2$. By associating with the fact that the number of particles at a distance between $r$ and $r + dr$ is $\rho g(r)4\pi r^2dr$, we can calculate the elastic energies of cooperative particles outside the cage to determine effects of their collective motions. The elastic barrier, $F_e$, is
\begin{eqnarray} 
F_{e} = 4\pi\rho\int_{r_{cage}}^{\infty}dr r^2 g(r)K_0\frac{u^2(r)}{2}. 
\label{eq:5}
\end{eqnarray}
For $r \geq r_{cage}$, $g(r)\approx 1$. The calculations allow us to determine contributions of nearest-neighbor interactions and collective rearrangement to the activated relaxation of a particle.

Due to chemical and biological complexities, conformational configuration, and chain connectivity, local and non-local dynamics is non-universally coupled. In our recent work \cite{9}, an adjustable parameter $a_c$ is introduced to scale the collective elastic barrier as $F_e \rightarrow a_cF_e$. The treatment has simultaneously provided quantitatively good agreements between theory and experiment in both the dynamic fragility and temperature dependence of structural relaxation time for 22 amorphous drugs and polymers \cite{9}. According to Kramer's theory, the structural (alpha) relaxation time defined by the mean time for a particle to diffuse from its particle cage is
\begin{eqnarray}
\frac{\tau_\alpha}{\tau_s} = 1+ \frac{2\pi}{\sqrt{K_0K_B}}\frac{k_BT}{d^2}\exp\left(\frac{F_B+a_cF_e}{k_BT} \right),
\label{eq:6}
\end{eqnarray}
where $K_B$=$\left|\partial^2 F_{dyn}(r)/\partial r^2\right|_{r=r_B}$ is absolute curvatures at the barrier position and $\tau_s$ is a short time scale of relaxation. The explicit expression of $\tau_s$ is \cite{6,7}
\begin{eqnarray}
\tau_s=g^2(d)\tau_E\left[1+\frac{1}{36\pi\Phi}\int_0^{\infty}dq\frac{q^2(S(q)-1)^2}{S(q)+b(q)} \right],
\label{eq:6-1}
\end{eqnarray}
where $\tau_E$ is the Enskog time scale,  $b(q)=1/\left[1-j_0(q)+2j_2(q)\right]$, and $j_n(x)$ is the spherical Bessel function of order $n$. In various works \cite{2,6,7,42,9} of thermal liquids, polymers and amorphous drugs, $\tau_E \approx 10^{-13}$ s.

To compare our hard-sphere calculations with experiment, a density-to-temperature conversion (thermal mapping) is required. The initial thermal mapping proposed by Schweizer \cite{10} is
\begin{eqnarray}
S_0^{HS}(\Phi)&=&\frac{(1-\Phi)^4}{(1+2\Phi)^2}\equiv S_0^{exp}(T,P),
\label{eq:Cmapping}
\end{eqnarray}
where $S_0$ is the isothermal compressibility. Clearly, this mapping requires experimental equation-of-state (EOS) data. The superscripts $HS$ and $exp$ correspond to hard sphere and experiment, respectively. Although this mapping has successfully provided both qualitative and quantitative descriptions for $\tau_\alpha(T)$ for 17 polymers and thermal liquids \cite{2,6,7,8,10}, the EOS data is unknown for our three polymers presented in next sections. 

Thus, we employ another thermal mapping \cite{42,9,50} constructed from the thermal expansion process of materials. During a heating process, the number of molecules remains unchanged while the volume of material increases linearly. This analysis leads to $\rho \approx \rho_0\left[1-\beta\left(T-T_{0}\right)\right]$ \cite{42,9,50}. Here $\beta$ is the volume thermal expansion coefficient, and $\rho_0$ and $T_{0}$ are the initial number density and temperature, respectively. From this, we can convert from a volume fraction to temperature of experimental material via 
\begin{eqnarray}
T \approx T_0 - \frac{\Phi - \Phi_0}{\beta\Phi_0}.
\label{eq:7}
\end{eqnarray} 
For most organic materials and amorphous drugs (22 materials) \cite{42,9,50}, $\beta \approx 12\times 10^{-4}$ $K^{-1}$. This value is consistent with Schweizer's the original mapping \cite{2}. $\Phi_0 \approx 0.5$ is the characteristic volume fraction estimated in our prior works \cite{42,9,50}. The parameter $T_0$ captures material-specific details such as molar mass and particle size. This density-to-temperature conversion has been used in the cooperative-string model for supercooled dynamics \cite{51}. In our calculations, the parameters $T_0$ and $a_c$ are tuned to obtain the best quantitative agreement between theoretical and experimental temperature dependence of structural relaxation times.

\subsection{Ultra-local limit}
\begin{figure}[htp]
\center
\includegraphics[width=9cm]{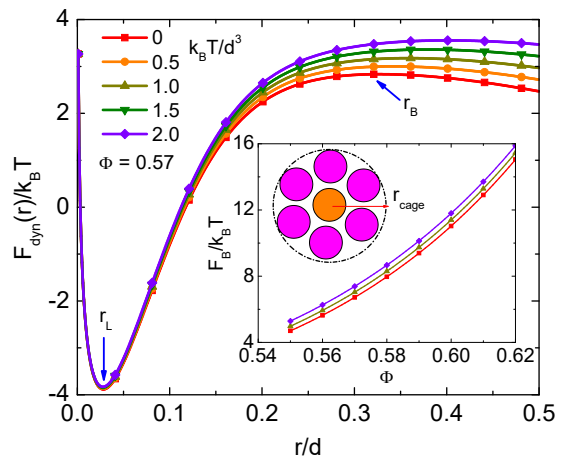}
\caption{\label{fig:3}(Color online) Dynamic free energy as a function of reduced particle displacement for a hard sphere fluid of packing fraction $\Phi = 0.57$ at several pressures in unit of $k_BT/d^3$. The inset shows a growth of the barrier height with $\Phi$ at $p = 0, 1,$ and 2 $k_BT/d^3$.}
\end{figure}

Figure \ref{fig:3} shows an example dynamic free energy for $\Phi = 0.57$ at different pressures in unit of $k_BT/d^3$ and defines key length and energy scales. The localization length is nearly insensitive to compression. Meanwhile, the barrier position increases and the local barrier height is raised with increasing the applied pressure. The result implies that the compression induces more constraint to the local dynamics of the tagged particle. 

When the local barrier is beyond a few  $k_BT$, much insight for key length scales of the dynamic free energy has been gained using the approximate "ultra-local" analytic analysis. In the ultra-local limit, since $r_L/d \ll 1$, high wavevectors are dominant in calculations of $F_{dyn}(r)$. We can ignore the wavevector integral below a cutoff $q_c$, and exploit $C(q)=-4\pi d^3 g(d)\cfrac{\cos(qd)}{(qd)^2}$ in the exact PY theory for $q \geq q_c$ \cite{1,11,12} and $S(q)\approx 1$.

Combining the analytical expression of $C(q)$ and $S(q)\approx 1$ with $\left[\partial F_{dyn}(r)/\partial r\right]_{r=r_L}=0$ gives a self-consistent equation for the localization length and barrier position

\begin{eqnarray}
\frac{9d^2}{r_{L,B}^2}-\frac{3P}{k_BT/d^3}\frac{d}{r_{L,B}}&\approx& \frac{24\Phi g^2(d)}{\pi}\int_{q_c}^{\infty}dq e^{-q^2r_{L,B}^2/3}\nonumber\\
&\approx& \frac{12\Phi g^2(d)}{\pi}\frac{\sqrt{3\pi}d}{r_{L,B}}erfc\left(\frac{q_cr_{L,B}}{\sqrt{3}}\right).\nonumber\\
\label{eq:12}
\end{eqnarray}

Now, since $q_cr_L/\sqrt{3} \ll 1$, one obtains
\begin{eqnarray}
r_L\equiv r_L(P)=\frac{r_L(P=0)}{1+\sqrt{\cfrac{\pi}{3}}\cfrac{P}{k_BT/d^3}\cfrac{1}{4g^2(d)\Phi}},
\label{eq:13}
\end{eqnarray}
where $r_L(P=0)=\cfrac{\sqrt{3\pi} d}{4g^2(d)\Phi}$ is the localization length at $P = 0$ or ambient pressure \cite{11,12}. Equation (\ref{eq:13}) quantitatively reveals how the external pressure restricts molecular motions. The localization length is reduced with increasing the compression. In addition, the Percus-Yevick (PY) theory for the contact number \cite{1} gives $g(d)=\cfrac{(1+\Phi/2)}{(1-\Phi)^2}$. Thus, $4\pi g^2(d)\Phi \approx 110$ for $\Phi=0.57$ is much larger than the considered values of $P/(k_BT/d^3)$. This finding explains why $r_L(P)$ is nearly unchanged as seen in Fig. \ref{fig:3}.

When $q_cr_B/\sqrt{3}$ is sufficiently large, one can use $\ce{erfc}(x)\approx e^{-x^2}/(\sqrt{\pi}x)$ to approximate $r_B$ in Eq. (\ref{eq:12}) and then obtain
\begin{eqnarray}
\frac{P}{k_BT/d^3}\frac{r_B}{d}&\approx& 3 - \frac{12\Phi g^2(d)}{\pi q_cd}\exp\left(-\frac{q_c^2r_B^2}{3} \right).
\label{eq:14}
\end{eqnarray}
The analytic form in Eq. (\ref{eq:14}) qualitatively indicates an increase of $r_B$ with increasing pressure as observed in Fig. \ref{fig:3}. Since prior works \cite{11,12} shows very poor quantitative accuracy of Eq. (\ref{eq:14}) compared to the numerical predictions at ambient pressure ($P\approx 0$), the deviation is expected to be large at elevated pressures. Thus, we do not show the corresponding curves.

The local barrier height $F_B$ in the ultra-local limit \cite{11,12} can be analytically calculated as
\begin{widetext}
\begin{eqnarray}
\frac{F_B}{k_BT} &=& -3\ln\frac{r_B}{r_L}-\frac{12\Phi g^2(d)}{\pi d} \int_{q_c}^\infty\frac{dq}{q^2}\left[e^{-q^2r_B^2/3}-e^{-q^2r_L^2/3} \right]\nonumber\\
&=& -3\ln\frac{r_B}{r_L}+\frac{12\Phi g^2(d)}{\sqrt{\pi}q_cd}\left[ \frac{q_cr_B}{\sqrt{3}}erfc\left(\frac{q_cr_B}{\sqrt{3}}\right)+\frac{e^{-q^2r_L^2/3}-e^{-q^2r_B^2/3}}{\sqrt{\pi}}\right].
\label{eq:16}
\end{eqnarray}
\end{widetext}
Clearly, the growth of $r_B$ with pressure is faster than that of $\ln(r_B)$ and it leads to the pressure-induced rise of $F_B$. At a given compression condition, we find that $F_B$ increases linearly with $\Phi g^2(d)= \cfrac{\Phi(1+\Phi/2)^2}{(1-\Phi)^4}$. Thus, $F_B$ grows with $\Phi$. The findings are consistent with numerical results shown in the inset of Fig. \ref{fig:3}. This analysis also reveals that adding the pressure term to the dynamic free energy as written in Eq. (\ref{eq:pressure}) exhibits the same manner as using Eq. (\ref{eq:2}) for hard-sphere fluids at higher effective volume fractions.

In addition, based on analysis in prior works \cite{11,12}, one can also perform an the dynamic shear modulus in the ultra-local limit as
\begin{eqnarray}
G(P) &=& \frac{9\Phi k_BT}{5\pi d r_L^2(P)}\nonumber\\
&=&\frac{9\Phi k_BT}{5\pi d r_L^2(P=0)}\left(1+\frac{P\sqrt{\pi/3}}{k_BT/d^3}\cfrac{1}{4g^2(d)\Phi}\right)^2\nonumber\\
&=& G(P=0)\left(1+\frac{P\sqrt{\pi/3}}{k_BT/d^3}\cfrac{1}{4g^2(d)\Phi}\right)^2.
\label{eq:15}
\end{eqnarray}
Equation (\ref{eq:15}) shows that $G(P)$ hardly changes with the applied pressure.
\section{Experimental Section}
\subsection{Materials}
\begin{figure}[htp]
\center
\includegraphics[width=8.5cm]{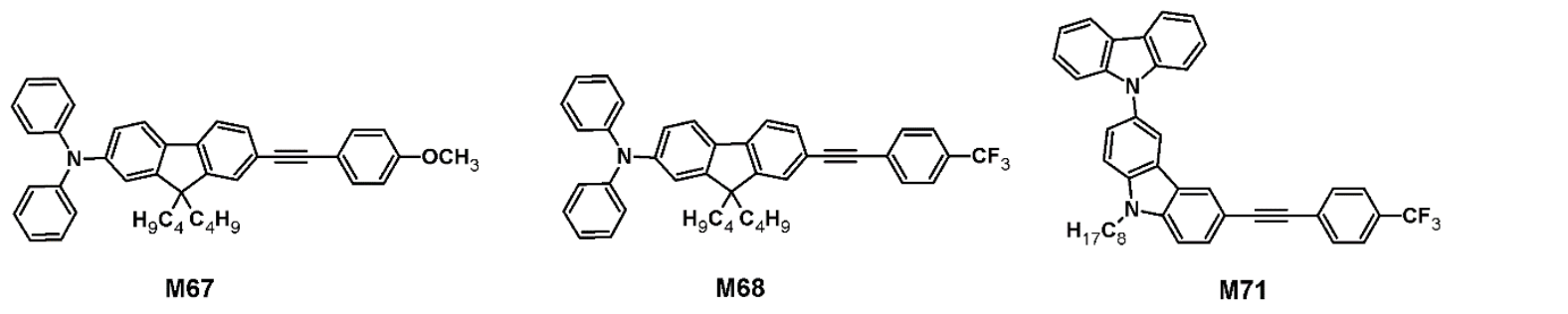}
\caption{\label{fig:1}(Color online) Chemical structures of tested compounds.}
\end{figure}

The experiments were performed on three rigid and non-polymeric supramolecules. Two of the tested samples are planar, linear, and their chemical structure (shown in Figure \ref{fig:1}) differs only in the end of the group (the diphenylamine-fluorene moiety is the same). In the material referred to M67 has the metoxy -OCH3 group is the end, while in sample named M68 the end of the group is the -CF3 moiety. The third material, entitled M71, enclose other motif (i.e. carbazole-carbazole group) than M67 and M68, which lead to deflection of the chemical structure. All of tested samples were synthesized by Sonogashira coupling reaction between 4-iodoanisole (M67) or 4-iodobenzotrifluoride (M68 and M71) and ethynyl derivative of diphenylamine-fluorene motif (M67 and M68) or ethynyl derivative of carbazole-carbazole moiety. The obtained compounds were purified by column chromatography, giving 99 $\%$ purity of the samples.

%\section{Experimental Method}
\subsection{Dielectric spectroscopy at ambient pressure}
The isobaric dielectric measurements at ambient pressure were carried out using Novo-Control GmbH alpha impedance analyzer in the frequency range from $10^{-2}$ to $10^6$ Hz at various temperature conditions (329-353 K for M67, 326-371 K for M68, and for M71 320-386 K). The temperature was controlled by Quatro temperature controller using a nitrogen gas cryostat with temperature stability better than 0.1 K. The tested sample was placed between two stainless steel electrodes of a capacitor (20 mm diameter) with a fixed gap between electrodes (0.1 mm) provided by fused silica spacer fibers. The dielectric measurements of M67 and M68 were performed after the vitrification by fast cooling from melting point (430, and 425 K, respectively), while M71 was measured during slow cooling from 386 K.
\subsection{Dielectric spectroscopy at elevated pressure}
The isothermal dielectric measurements at elevated pressure were performed utilizing a high-pressure system with an MP5 micropump (Unipress) and an alpha impedance analyzer (Novocontrol GmbH). The pressure was controlled with an accuracy better than 1 MPa by an automatic pressure pump, the silicone oil was used as a pressure-transmitting fluid. The sample cell was the same as used during the measurements at ambient pressure (15 mm diameter of the capacitor, 0.1 distance between electrodes provided by Teflon spacer). To avoid contact between sample and pressure-transmitting fluid, the capacitor was placed in a Teflon ring and additionally wrapped by the Teflon tape. The temperature was controlled by Weiss Umwelttechnik GmbH fridge with the precision better than 0.1 K. The measurements were performed at 347 K (5-45 MPa) for M67, 338 K (0.1 to 28 MPa) for M68 and 338 K (0.1 to 28 MPa) for M71.

%\section{Dielectric spectroscopy data analysis}

\section{Results and Discussion}
\begin{figure}[htp]
\center
\includegraphics[width=9cm]{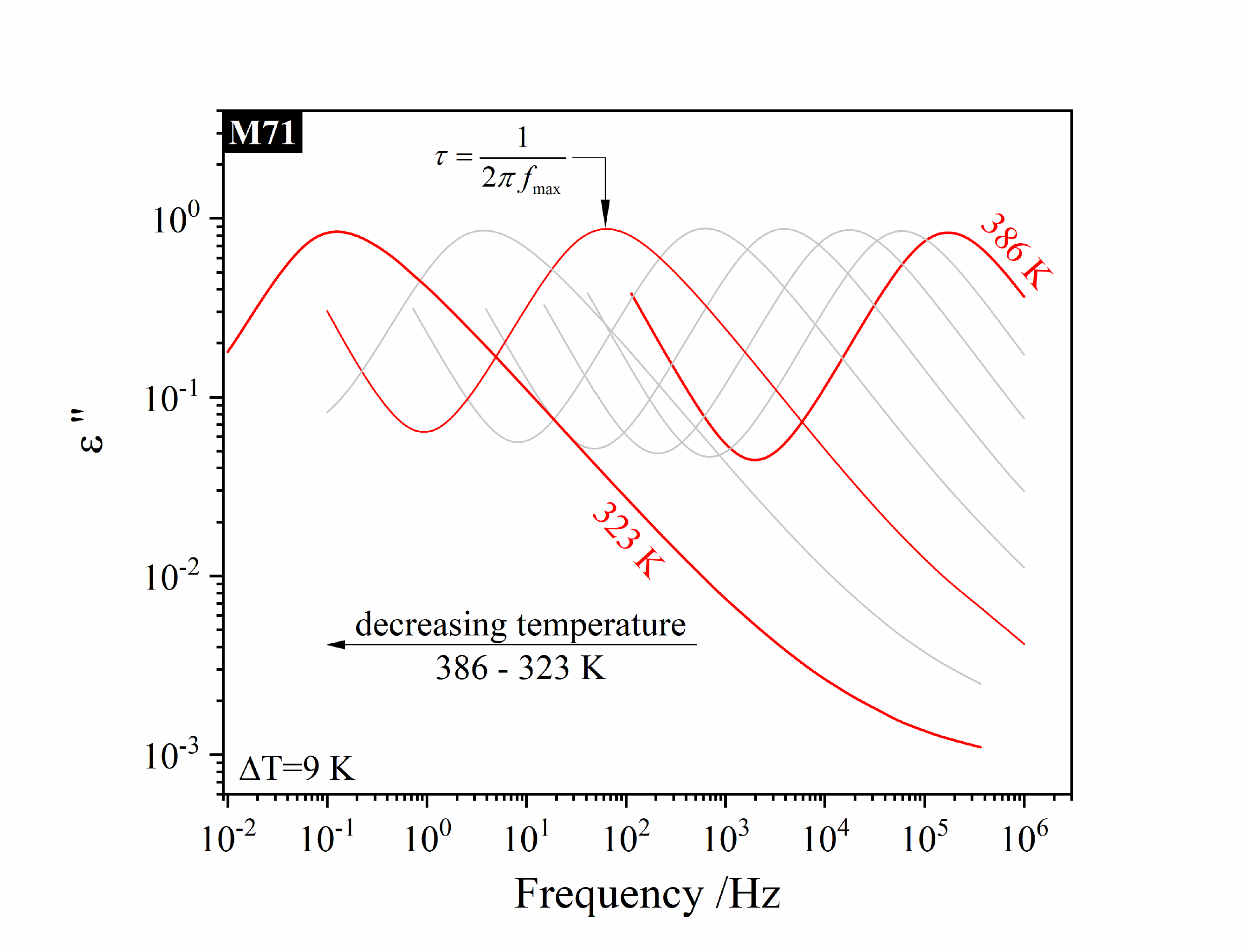}
\caption{\label{fig:2}(Color online) The dielectric loss spectra of M71 measured above glass transition temperature at ambient pressure.}
\end{figure}

\begin{figure}[htp]
\center
\includegraphics[width=9cm]{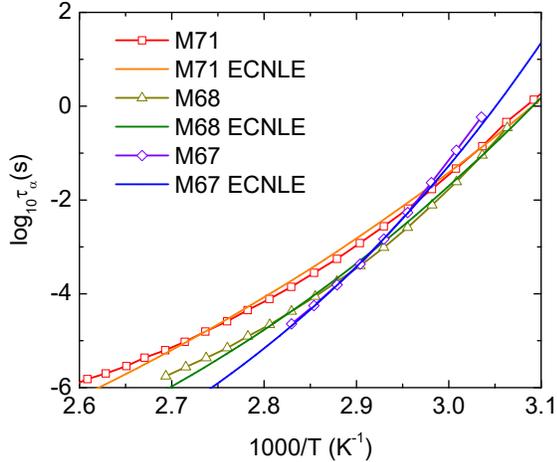}
\caption{\label{fig:4}(Color online) Temperature dependence of structural relaxation time of M67, M68, and M71 under ambient pressure ($P \approx 0$). Open points are experimental data and solid curves correspond to our ECNLE calculations.}
\end{figure}

\begin{figure}[htp]
\center
\includegraphics[width=9cm]{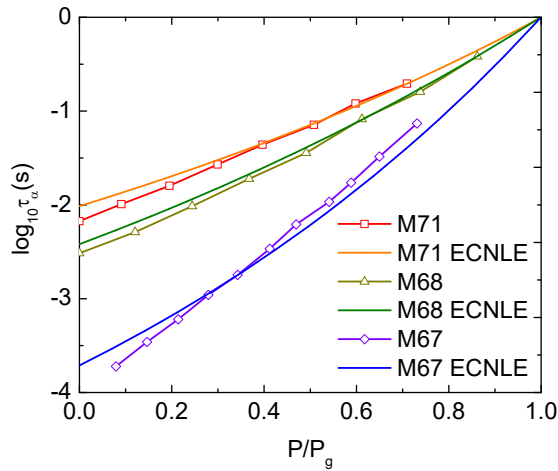}
\caption{\label{fig:5}(Color online) Logarithm of structural relaxation time of M67 at 347 $K$, M68 at 338 $K$, and M71 at 338 $K$  versus pressure normalized by $P_g$, which is defined by $\tau_\alpha(P_g)=1$s. Open points are experimental data and solid curves correspond to our ECNLE calculations.}
\end{figure}

Representative dielectric spectra measured for M71 above the glass transition temperature are presented in Fig. \ref{fig:2}. As can be seen, the structural relaxation process and dc-conductivity (on the low-frequency flank of the $\alpha$-process) move towards lower frequencies with decreasing temperature (or with squeezing at isothermal condition). From analysis of the dielectric loss peak we  obtained the relaxation time, $\tau_\alpha$, using the following definition: $\tau_\alpha=1/2\pi f_{max}$ , where $f_{max}$ is the maximum frequency of the structural relaxation peak. The $\log\tau_\alpha$ as a function of (i) inverse of temperature is presented in Fig. \ref{fig:4}, while (ii) $\log\tau_\alpha$ as a function of $P/P_g$ is depicted in Fig \ref{fig:5}. 

Figure \ref{fig:4} shows theoretical and experimental $\log_{10}\tau_\alpha$ of M67, M68, and M71 under atmospheric pressure ($P \approx 0$) as a function of $1000/T$. We use Eqs. (\ref{eq:6}), (\ref{eq:6-1}), and (\ref{eq:7}) to calculate the temperature dependence of $\tau_\alpha$. To obtain the quantitatively good accordance, we use $T_0=465$ $K$ and $a_c=4$ for M67, $T_0=499$ $K$ and $a_c=1$ for M68, and $T_0=524$ $K$ and $a_c=0.36$ for M71. Different chemical end groups cause the different relative importance of the collective elastic distortion and give various values of $a_c$. Overall, the ENCLE calculations agree quantitatively well with experimental data.

Under high compression effects, motion of particles has more constraint and the relaxation process is significantly slowed down. From the previous section, we know that the barrier height $F_B$ and jump distance $\Delta r=r_B-r_L$ are increased with a pressure rise. Thus, the collective barrier $F_e \sim K_0\Delta r^4$ also grows. For simplification, we assume that the correlation between local and collective molecular dynamics in substances does not change when applying pressure. In addition, the thermal expansion coefficient $\beta$ and the characteristic temperature $T_0$ are supposed to remain unchanged with pressure. The assumption allows us to calculate the pressure dependence of structural relaxation time. Since pressure entering to the dynamic free energy in Eq. (\ref{eq:pressure}) is in unit of $k_BT/d^3$, our numerical results can be compared to experimental data without introducing additional parameters by the pressure normalization.

Theoretical calculations and experimental data for $\log_{10}\tau_\alpha$ versus normalized pressure of our three materials in an isothermal condition are contrasted in Fig. \ref{fig:5}. At a fixed temperature, we use Eq. (\ref{eq:7}) to map from temperature to a packing fraction of the effective hard-sphere fluid in ECNLE calculations. Then, the pressure dependence of physical quantities for local dynamics and the alpha relaxation time are calculated using Eq. (\ref{eq:pressure}) when varying pressure. We define the glass transition pressure $P_g$ at $\tau_\alpha(P_g)=1s$ to normalize pressure. One observes a quantitatively good accordance between theory and experiment shown in Fig. \ref{fig:5}. This agreement suggests that our simple assumption of ignoring effects of chemical and biological structures seems plausible. We do not need to consider steric repulsion between molecules since the hard-sphere models are still applicable during compression. However, this simplicity may cause deviation between theory and experiment. Numerical results in Fig.\ref{fig:5} also reveal that our extended ECNLE theory is a predictive approach to investigate effects of pressure when only knowing parameters $T_0$ and $a_c$ from molecular mobility at ambient conditions. %The activated events over a wide range of compression and timescale are well-described using the ECNLE theory.

\begin{figure}[htp]
\center
\includegraphics[width=9cm]{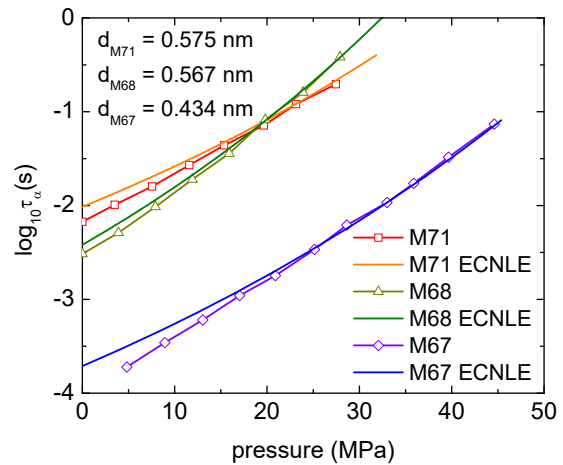}
\caption{\label{fig:6}(Color online) Logarithm of structural relaxation time of M67 at 347 $K$, M68 at 338 $K$, and M71 at 338 $K$  versus pressure in unit of MPa. Open points are experimental data and solid curves correspond to our ECNLE calculations.}
\end{figure}

To compare with experiment in real unit of pressure (MPa), we establish an equality between the theoretical and experiment $P_g$ to calculate the particle diameter. Results are $d = 0.434$ nm for M67, $d = 0.567$ nm for M68, and $d = 0.575$ nm for M71, respectively. Experimental data and theoretical calculations for the pressure dependence of $\tau_\alpha$ of our three pure amorphous materials in isothermal processes are shown in Figure \ref{fig:6}. One can see better quantitative consistency between theory and experiment than in Fig. \ref{fig:5} since $d$ is fixed and calculated at $P = P_g$. At high-pressure regime, molecules are incompressible while at low pressures (and/or ambient condition), molecules are internally relaxed and their volume becomes relatively larger. The curves of ECNLE calculations are slightly above those of experimental data.  The theory-experiment deviation becomes more important at low compression. 

\begin{figure}[htp]
\center
\includegraphics[width=9cm]{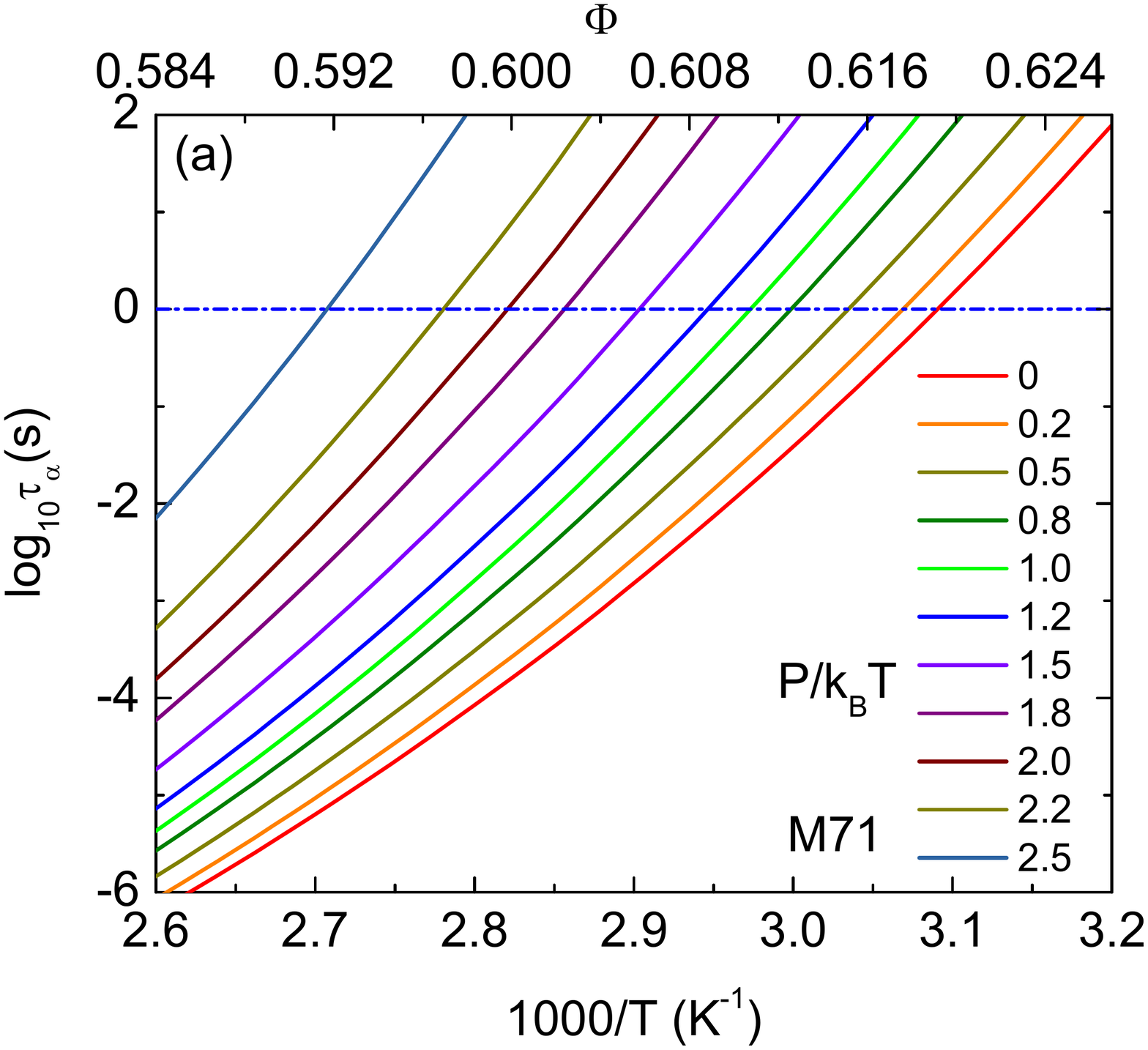}
\includegraphics[width=9cm]{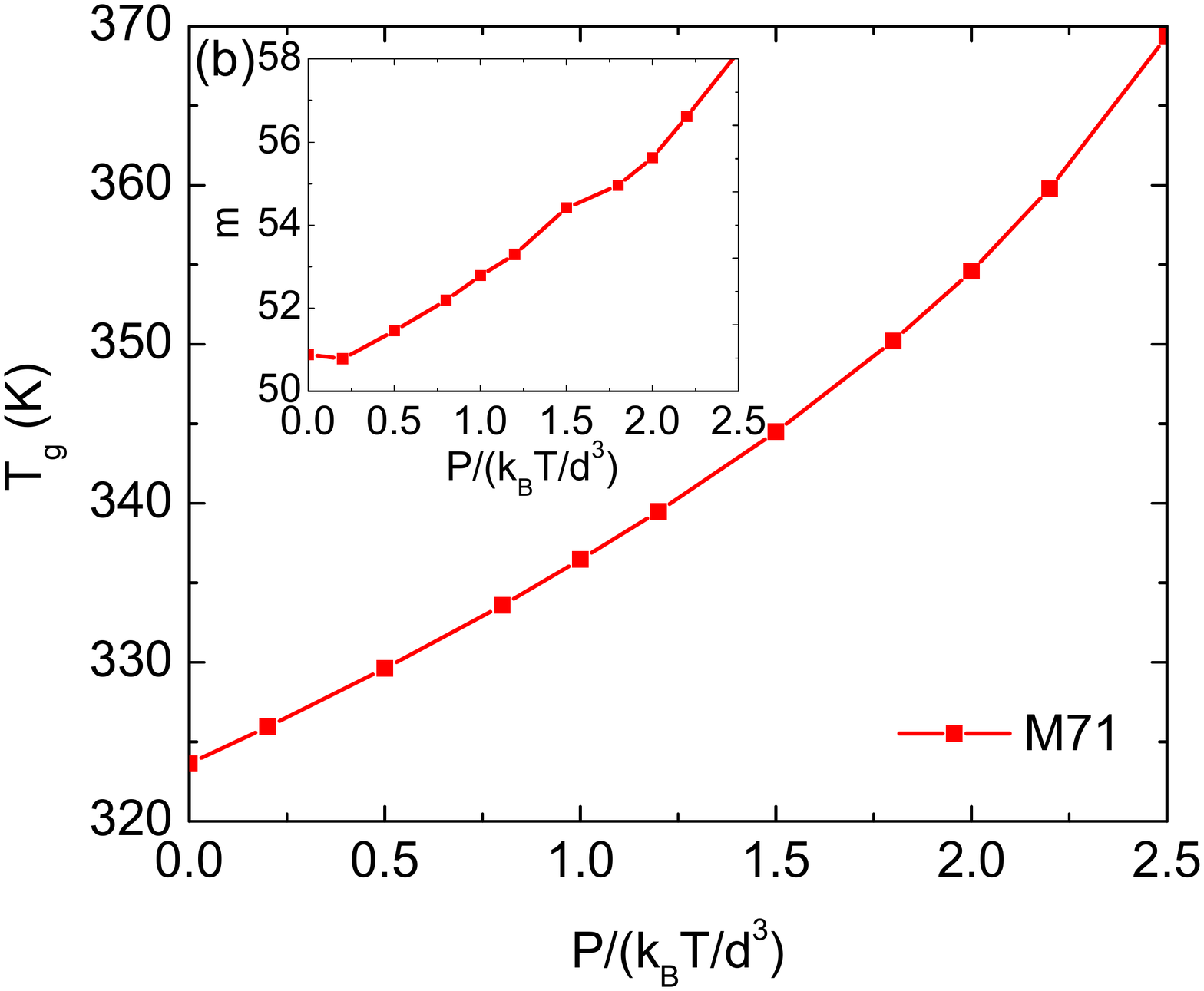}
\caption{\label{fig:7}(Color online) (a) Logarithm of structural relaxation time of M71 at different external pressures. A horizontal blue dashed line indicates a vitrification time scale
criterion of 1 $s$. (b) The pressure dependence of the glass transition temperature of M71. The inset shows the theoretical fragility plotted versus external pressures in unit of $k_BT/d^3$.}
\end{figure}

Obviously, there is no universal way to determine $d$. If the diameter $d$ is calculated at a low pressure regime, the behavior is reversed and theoretical predictions deviate from experiment at high pressures. These results clearly indicate that the external pressure not only reduces the free volume, but also change the molecular size. All factors change the packing fraction $\Phi$. In Fig. \ref{fig:7}a, we show the temperature or density dependence of $\tau_\alpha$ for a representative material (M71) under various pressure conditions. Increasing the packing fraction $\Phi$ and compression slows down the molecular dynamics in the same manner. The shrinking-down process of molecules under large compression can be quantified by tuning the value of $d$ to obtain the best quantitative fit between theoretical and experimental $\log_{10}\tau_\alpha(P)$.

Based on theoretical calculations in Fig. \ref{fig:7}a, one can determine $T_g(P)$ defined as $\tau_\alpha(T_g)=1$ s and the dynamic fragility of M71
\begin{eqnarray}
m = \left. \frac{\partial\log_{10}(\tau_\alpha)}{\partial(T_g/T)}\right |_{T=T_g}.
\label{eq:fragility}
\end{eqnarray} 
Numerical results are shown in Fig. \ref{fig:7}b. Generically, both $T_g$ and $m$ increase with compression. It means this glass former becomes more fragile at elevated pressure. In the ECNLE theory, the higher fragility corresponds to more collective elasticity or greater effects of collective motions on the glass transition \cite{8,9}. This finding is consistent with prior simulations \cite{71,73} and experiments \cite{72,74}. We can explain this behavior using a nontrivial correlation among the cooling rate ($h$), glass transition temperature, and dynamic fragility \cite{9}
\begin{eqnarray}
h\tau_\alpha(T_g) = \frac{T_g}{m\ln(10)}.
\label{eq:18}
\end{eqnarray}
Since $h\tau_\alpha(T_g)$ is a constant, $m$ monotonically vary with $T_g$. Consequently, at a fixed temperature, the pressure-induced slowing down of the relaxation time shifts $T_g$ towards a larger value and causes an increase of $m$. We emphasize that this analysis can be changed if glass-forming liquids have strong electrostatic interactions and chemical/biological complexities.

\section{Conclusions}
We have developed the ECNLE theory of bulk relaxation to capture the pressure effects on the glass transition of glass-forming liquids. Amorphous materials are described as a hard sphere fluid. Under compression condition, a mechanical work done by the pressure acting on a tagged particle modifies its the dynamic free energy. The free energy profile provides the pressure dependence of key physical quantities of the local dynamics by only considering nearest-neighbor interactions. The localization length is slightly reduced with increasing pressure, while the barrier position and local-barrier height grows. These variations in the ultra-local limit (high densities or low temperatures) have been analytically analyzed. Our calculations indicate that further restrictions apply to the local dynamics. It leads to a significantly slowing-down of molecular mobility when applying pressure. The validity of our theoretical approach has been supported by dielectric spectroscopy experiments. We measured the dielectric loss spectra of three different materials to determine the alpha structural relaxation time at ambient and elevated pressures over a wide range of temperature. Our theoretical temperature and pressure dependence of the structural relaxation time quantitatively agree with experimental data. %\textcolor{red}{These calculations also reveal effects of pressure-induce variation of the molecular size on the glass transition}. This problem is under study.

\begin{acknowledgments}
This work was supported by JSPS KAKENHI Grant Numbers JP19F18322 and JP18H01154. M. Paluch is deeply grateful for the financial support by the National Science Centre within the framework of the Maestro10 project (grant no UMO- 2018/30/A/ST3/00323). This research was funded by the Vietnam National Foundation for Science and Technology Development (NAFOSTED) under grant number 103.01-2019.318.
\end{acknowledgments}

\section*{Conflicts of interest}
There are no conflicts to declare.

\end{document}